\documentclass
[showpacs,showkeys,nofootinbib,groupedaddress,titlepage,preprint,12pt,prd,eqsecnum]{revtex4}%
\usepackage{amsfonts}
\usepackage{amsmath}
\usepackage{amssymb}
\usepackage{graphicx}%
\begin{document}
\title{Moduli Vacuum Bubbles Produced by Evaporating Black Holes\\ }
\author{J.R. Morris}
\affiliation{Physics Dept., Indiana University Northwest, 3400 Broadway, Gary, Indiana
46408, USA}
\email{jmorris@iun.edu}

\begin{abstract}
We consider a model with a toroidally compactified extra dimension giving rise
to a temperature-dependent 4d effective potential with one-loop contributions
due to the Casimir effect, along with a 5d cosmological contant. The forms of
the effective potential at low and high temperatures indicates a possibility
for the formation of a domain wall bubble, formed by the modulus scalar field,
surrounding an evaporating black hole. This is viewed as an example of a
recently proposed black hole vacuum bubble arising from matter-sourced moduli
fields in the vicinity of an evaporating black hole [D. Green, E. Silverstein,
and D. Starr, Phys. Rev. D74, 024004 (2006)]. The black hole bubble can be
highly opaque to lower energy particles and photons, and thereby entrap them
within. For high temperature black holes, there may also be a
symmetry-breaking black hole bubble of false vacuum of the type previously
conjectured by Moss [I.G. Moss, Phys. Rev. D32,1333 (1985)], tending to
reflect low energy particles from its wall. A double bubble composed of these
two different types of bubble may form around the black hole, altering the
hole's emission spectrum that reaches outside observers. Smaller mass black
holes that have already evaporated away could have left vacuum bubbles behind
that contribute to the dark matter.

\end{abstract}

\maketitle

\section{Introduction}

A black hole freely radiating into empty space through the Hawking
process\cite{Hawking} has an associated temperature $T_{h}=1/(8\pi GM_{h})$
where $M_{h}$ is the mass of the black hole. The evaporation rate is inversely
proportional to the mass, $\dot{M}_{h}\propto-1/(GM_{h})^{2}$ so that
primordial black holes (PBHs) created soon after the big bang with masses
$M_{h}\lesssim10^{12}$ kg may have already evaporated away. PBHs with mass
$\sim10^{12}$ kg and size of $\sim$ 1 fm may still be present, and quite hot.
Moss\cite{Moss} has pointed out the possibility that a high temperature
symmetric phase can surround the black hole. Further away, the temperature
drops and symmetry is broken. These symmetric and broken symmetric phases are
separated by a domain wall which surrounds the black hole, i.e., the PBH lies
within a \textquotedblleft black hole bubble\textquotedblright. A particle may
have a mass $m$ outside the bubble, while inside the bubble -- in the
symmetric phase -- the particle mass vanishes by the Higgs mechanism. This
particle will be totally reflected from the inner wall of the bubble if its
energy is $E<m$, since it is energetically trapped. Trapped particles can help
stabilize the bubble against collapse, and there can be an approach to a state
of thermal equilibrium.

Recently, Green, Silverstein, and Starr\cite{GSS} have conjectured that a
different type of vacuum bubble, associated with scalar moduli fields sourced
by matter fields of compact objects, may be catalyzed by evaporating black
holes. A realization of this type of scenario leads to the possibility that
such \textquotedblleft moduli vacuum bubbles\textquotedblright\ may be end
products of the Hawking radiation of black holes. These moduli vacuum bubbles
can arise from the local effects of the matter sources coupling to one or more
of the scalar moduli fields in the effective low energy field theory.

Here we consider the possible effects of extra dimensions and the formation of
a modulus vacuum bubble near a hot PBH by employing a model with one extra
space dimension that is toroidally compactified, but in an inhomogeneous way.
If the extra dimension has an associated scale factor $B(x^{\mu})$, the
physical size of the compact dimension is $(2\pi R_{5})B(x^{\mu})$, where
$R_{5}$ is the radius of the compactified dimension. The 4d effective
potential that arises from the extra dimension is temperature-dependent, and
the equilibrium value of $B$ may be quite different in the high and low
temperature regions, so that the compactification is inhomogeneous. A modulus
field $\varphi(x^{\mu})\propto\ln B(x^{\mu})$ can then give rise to the
modulus bubble surrounding the PBH, with $B$ interpolating between two
different values on either side of the bubble wall. The bubble wall has a
thickness $\delta$ and separates the hot region near the PBH and the cold
asymptotic region. The presence of a domain wall with a variation in
$B(x^{\mu})$ over a distance $\delta$ can have dramatic effects on photons and
other particles, and generally leads to energy-dependent reflectivities. For
ordinary (nonmodulus) domain walls, reflection probabilities are relatively
large for lower particle energies $E\ll\delta^{-1}$, while transmission
probabilties become large for higher energies $E\gg\delta^{-1}$. (See, e.g.,
refs.\cite{VSbook,Everett,EDW1,EDW2}.) The same type of behavior has been
found for the case of domain walls associated with moduli\cite{DLM}.
Therefore, lower energy particles can become trapped, at least to some degree,
within a bubble and help to stabilize it against collapse.

If both types of bubble co-exist--i.e., a modulus vacuum bubble and a
nonmodulus false vacuum bubble from symmetry restoration at high temperature--
there will be a \textquotedblleft double bubble\textquotedblright\ containing
the PBH, interfering with the escape of various types of particles being
emitted from it. The Hawking radiation will be partially trapped within the
bubble, and to a distant observer an evaporating black hole may have a
transmitted energy spectrum that is somewhat different from what would be
expected from a black hole freely radiating into empty space, with the lower
frequency portions of the emission spectrum being suppressed. PBHs that have
already evaporated away may have left metastable vacuum bubbles behind that
contribute to the dark matter of the universe.

In section 2 the effective 4d Einstein frame effective theory is obtained from
the 5d theory. The form of the 4d effective potential $U$ \ (developed in the
appendices) for the scalar modulus containing contributions from one loop
quantum corrections at finite temperature, along with a cosmological constant,
is presented, and the low and high temperature limits of $U$ are examined. In
section 3 we consider black hole bubbles, both the (nonmodulus) symmetry
breaking type and the modulus type. A brief summary forms section 4. In
Appendix A the dependence of the effective potential $U$ (for a flat Einstein
frame) upon the Rubin-Roth action density and the cosmological constant is
shown. Asymptotic forms of $U$ are obtained in Appendix B.

\section{Effective 4d Theory}

\subsection{Metric and Effective Action}

We consider a 5d spacetime with a topology of $M_{4}\times S^{1}$ having one
toroidally compact extra dimension. The (mostly negative) 5d metric is
$\tilde{g}_{MN}$:%
\begin{equation}
ds^{2}=\tilde{g}_{MN}dx^{M}dx^{N}=\tilde{g}_{\mu\nu}dx^{\mu}dx^{\nu}%
-B^{2}dy^{2} \label{e1}%
\end{equation}
where $M,N=0,1,2,3,5$, $\mu,\nu=0,1,2,3,$ and $y=x^{5}$ is the coordinate of
the compact extra dimension, $0\leq y\leq2\pi R_{5}$. We allow $\tilde{g}%
_{\mu\nu}$ and $\sqrt{-\tilde{g}_{55}}=B$ to have a dependence on $x^{\mu}$,
but assume them to be independent of $y$. We also assume that $\tilde{g}%
_{\mu5}=0$. The scale factor $B(x^{\mu})$ can be related to a scalar (modulus)
field $\varphi(x^{\mu})$ and the circumference of the extra dimension
$L_{5}=(2\pi R_{5})\sqrt{-\tilde{g}_{55}}$ by%
\begin{equation}
B=e^{\sqrt{2/3}\kappa\varphi},~~~~~\varphi=\frac{1}{\kappa}\sqrt{\frac{3}{2}%
}\ln B,~~~~~L_{5}=(2\pi R_{5})B \label{e2}%
\end{equation}
where $\kappa=\sqrt{8\pi G}=\sqrt{8\pi}/M_{P}$, with $M_{P}=1/\sqrt{G}$ the
Planck mass. We can talk in terms of the scalar field $\varphi$, the scale
factor $B$, or the circumference of the extra dimension $L_{5}$
interchangeably. We will want to focus attention upon a 4d effective potential
$U$ (which can be regarded as a function of either $\varphi$, $B$, or $L_{5}$)
which originates from the Rubin-Roth potential describing one-loop quantum
corrections at finite temperature due to Casimir effects for bosons and
fermions\cite{RR1,RR2}, along with a contribution from a 5d cosmological
constant $\Lambda$. Although, by eq.(\ref{e2}), this effective potential will
take different functional forms when expressed in terms of $\varphi$ or $B$,
we will describe it simply as $U(\varphi)$, $U(B)$, or $U(L_{5})$ when
confusion is not likely to arise. At low temperatures, Blau and
Guendelman\cite{Blau-Guen} demonstrated that there are parameter ranges for
the effective potential $U$, allowing an inhomogeneous compactification to
occur, so that in the effective 4d theory the scalar field $\varphi(x^{\mu})$
can be associated with a domain wall that smoothly connects two vacuum states.
The resulting \textquotedblleft dimension bubbles\textquotedblright\ have
peculiar properties\cite{JM1,G-M,JM2} and can be stabilized by the entrapment
of massive particle modes and/or photons.

We start with a 5d action%
\begin{equation}%
\begin{array}
[c]{cc}%
S & =\frac{1}{2\kappa_{5}^{2}}\int d^{5}x\sqrt{\tilde{g}_{5}}\left\{
\tilde{R}_{5}-2\Lambda+2\kappa_{5}^{2}\mathcal{L}_{5}\right\} \\
& =\frac{1}{2\kappa^{2}}\int d^{4}x\sqrt{-\tilde{g}}B\left\{  \tilde{R}%
_{5}-2\Lambda+2\kappa^{2}\mathcal{L}\right\}
\end{array}
\label{e3}%
\end{equation}
where we have used the definitions $\kappa_{5}^{2}=8\pi G_{5}=(2\pi
R_{5})\kappa^{2}$, $\mathcal{L}=(2\pi R_{5})\mathcal{L}_{5}$, $\tilde{g}%
_{5}=\det\tilde{g}_{MN}$, and $\tilde{g}=\det\tilde{g}_{\mu\nu}$. Also,
$\tilde{R}_{5}=\tilde{g}^{MN}\tilde{R}_{MN}$ is the 5d Ricci scalar built from
$\tilde{g}_{MN}$. The 4d Jordan frame metric is $\tilde{g}_{\mu\nu}$, the
$\mu\nu$ part of $\tilde{g}_{MN}$. We define a 4d Einstein frame metric
$g_{\mu\nu}$ by $g_{\mu\nu}=B\tilde{g}_{\mu\nu}=e^{\sqrt{\frac{2}{3}}%
\kappa\varphi}\tilde{g}_{\mu\nu}=(L_{5}/2\pi R_{5})\tilde{g}_{\mu\nu}$. The
line element in eq.(\ref{e1}) then becomes%
\begin{equation}%
\begin{array}
[c]{ll}%
ds^{2} & =B^{-1}g_{\mu\nu}dx^{\mu}dx^{\nu}-B^{2}dy^{2}\\
& =e^{-\sqrt{\frac{2}{3}}\kappa\varphi}g_{\mu\nu}dx^{\mu}dx^{\nu}%
-e^{2\sqrt{\frac{2}{3}}\kappa\varphi}dy^{2}%
\end{array}
\label{e4}%
\end{equation}
Using (\ref{e3}) and (\ref{e4}), the 5d action is dimensionally reduced to an
effective 4d Einstein frame action%
\begin{equation}%
\begin{array}
[c]{cc}%
S & =\int d^{4}x\sqrt{-g}\left\{  \frac{1}{2\kappa^{2}}R+\frac{1}{2}%
(\nabla\varphi)^{2}+e^{-\sqrt{\frac{2}{3}}\kappa\varphi}[\mathcal{L}%
-\Lambda/\kappa^{2}]\right\} \\
& =\int d^{4}x\sqrt{-g}\left\{  \frac{1}{2\kappa^{2}}R+\frac{3}{4\kappa
^{2}B^{2}}(\nabla B)^{2}+B^{-1}[\mathcal{L}-\Lambda/\kappa^{2}]\right\}
\end{array}
\label{e5}%
\end{equation}
where $R=g^{\mu\nu}R_{\mu\nu}$ is the 4d Einstein frame Ricci scalar built
from $g_{\mu\nu}$.

From (\ref{e5}) we see that in the effective 4d theory the extra dimensional
scale factor enters as a scalar field and that there is a 4d effective
Lagrangian $\mathcal{L}_{4}=B^{-1}\mathcal{L}$ produced by the Lagrangian
$\mathcal{L}$. A total effective 4d Lagrangian can therefore be written as%
\begin{equation}
\mathcal{L}_{eff}=\frac{1}{2\kappa^{2}}R+\frac{1}{2}(\nabla\varphi
)^{2}-U(\varphi)+\mathcal{L}_{4} \label{e6}%
\end{equation}
where $U(\varphi)$ is an effective potential that is constructed from one-loop
quantum corrections for fermions and bosons at finite temperature (Rubin-Roth
potential), along with the cosmological constant term.

The temperature-dependent effective potential $U$ can be written in terms of
$L_{5}$ and $\beta$, and its basic structure (expressed in a flat Einstein
frame spacetime background) showing its dependence upon the Rubin-Roth
potential $\tilde{\Gamma}$ and the cosmological constant $\Lambda$ is given by
(\ref{ap16}) in Appendix A. The result is%
\begin{equation}
U(L_{5},\beta)=(2\pi R_{5})^{2}\dfrac{\tilde{\Gamma}(L_{5},\beta)}{\beta
L_{5}^{2}}+(2\pi R_{5})\dfrac{\Lambda/\kappa^{2}}{L_{5}} \label{b7}%
\end{equation}
The asymptotic forms of $\tilde{\Gamma}$ for low and high temperatures are
given in Tables 1 and 2 (taken from Table 1 in ref.\cite{RR2}). Using these we
can express $U(L_{5},\beta)$ for high and low $\beta$ limits. These limiting
forms (see eqs. (\ref{ap25}) and (\ref{ap26}) in Appendix B) are given by%
\begin{equation}
U\approx\left\{
\begin{array}
[c]{ll}%
\lbrack(N_{f}+n_{f})-(N_{b}+n_{b})](2\pi R_{5})^{2}\dfrac{3\zeta(5)}{4\pi^{2}%
}\dfrac{1}{L_{5}^{6}}+\dfrac{(2\pi R_{5})\Lambda}{\kappa^{2}L_{5}},\smallskip
& L_{5}\ll1/M\\
(n_{f}-n_{b})(2\pi R_{5})^{2}\frac{M^{2}}{4\pi^{2}}e^{-ML_{5}}\dfrac{1}%
{L_{5}^{4}}+\dfrac{(2\pi R_{5})\Lambda}{\kappa^{2}L_{5}}, & L_{5}\gg1/M
\end{array}
\right\}  \text{\ (low }T\text{)} \label{b8}%
\end{equation}
for the high $\beta$ limit, where $M$ is a typical particle mass parameter,
and%
\begin{equation}
U\approx\left[  (N_{b}+\frac{15}{16}N_{f})(2\pi R_{5})^{2}\frac{3\zeta
(5)}{4\pi^{2}}\right]  \frac{1}{\beta_{1}^{5}}\left\{
\begin{array}
[c]{cc}%
\dfrac{(N_{f}-N_{b})}{(N_{b}+\frac{15}{16}N_{f})}\dfrac{\beta_{1}^{5}}%
{L_{5}^{6}}+\dfrac{1}{L_{5}},\smallskip & L_{5}\ll\beta\\
\left(  1-\dfrac{\beta_{1}^{5}}{\beta^{5}}\right)  \dfrac{1}{L_{5}}, &
L_{5}\gg\beta
\end{array}
\right\}  \ \ (\text{high }T) \label{b9}%
\end{equation}
for the low $\beta$ limit. Here, $n_{f(b)}=$ \# massive fermionic (bosonic)
modes, each of mass $\sim M$, and $N_{f(b)}=$ \# effectively massless modes,
with $N_{b}\geq5$ (5 graviton degrees of freedom). We assume $N_{f}>N_{b}$ in
(\ref{b9}) and $(N_{f}+n_{f})-(N_{b}+n_{b})>0$ in (\ref{b8}). The parameter
$\beta_{1}$ is defined by (see eq. (\ref{ap23}))%
\begin{equation}
T_{1}=\frac{1}{\beta_{1}}=\left\{  \frac{\Lambda/\kappa^{2}}{(N_{b}+\frac
{15}{16}N_{f})(2\pi R_{5})\frac{3\zeta(5)}{4\pi^{2}}}\right\}  ^{1/5}
\label{b10}%
\end{equation}
If we want to explicitly account for differences in particle masses in
(\ref{b8}), we can make the replacement%
\begin{equation}
n\frac{M^{2}}{4\pi^{2}}\frac{\beta}{L_{5}^{4}}e^{-ML_{5}}\rightarrow\sum
_{i}\left[  n^{(i)}\frac{M_{i}^{2}}{4\pi^{2}}\frac{\beta}{L_{5}^{4}}%
e^{-M_{i}L_{5}}\right]  \label{b11}%
\end{equation}
with the index $i$ running over the different species. The expression for $U$
in terms of the scalar field $\varphi$ is obtained by \ simply replacing
$L_{5}$ by $(2\pi R_{5})e^{\sqrt{\frac{2}{3}}\kappa\varphi}$.

We consider three distinct temperature ranges: a low temperature (high
$\beta=1/T$) regime where $T\ll M$, where $M$ is a typical particle mass; a
high temperature regime where $T\gg M$ but $T<T_{1}$; and a very high
temperature regime where $T\gg M$ and $T>T_{1}$.

The precise form of the effective potential depends upon the values of the
various parameters, such as the 5d cosmological constant $\Lambda$ and the
compactification radius $R_{5}$, which are not known, but we consider (as in
refs.\cite{Blau-Guen,JM1}) the interesting case for which the low temperature
potential $U$ \ has a local minimum at some value $L_{5}=L_{5,\min}$
($\varphi=\varphi_{\min}$), a local maximum at $L_{5}=L_{5,\max}>L_{5,\min}$,
and $U\rightarrow0$ as $L_{5}\rightarrow\infty$. There are then two low energy
states, at $L_{5,\min}$ and $L_{5}>L_{5,\max}$, separated by a potential
barrier. A scalar field domain wall solution $\varphi$ interpolating between
these two low temperature \textquotedblleft vacuum\textquotedblright\ states
forms the wall of a low temperature \textquotedblleft dimension
bubble\textquotedblright\cite{Blau-Guen,JM1}. A schematic depiction of
$U(L_{5},\beta)$ for low $T$ is sketched in Fig.1. (We will later argue that
the existence of a local minimum in the low temperature potential, though
required for the existence of a dimension bubble, is not required for the
existence of a black hole modulus bubble.)

One can actually distinguish between two different possible types of low
temperature dimension bubble\cite{JM2}. A type I bubble has a large value of
$L_{5}$ in its interior and $L_{5}\rightarrow L_{5,\min}$ outside the bubble.
A type II bubble has $L_{5}=L_{5,\min}$ in its interior and $L_{5}>L_{5,\max}$
outside. In each case the bubble forms because $U_{inside}>U_{outside}$,
causing the domain wall to bend and enclose a region of higher energy density.
Both types tend to entrap low energy photons having wavelengths greater than
the wall thickness. The photon and particle reflectivity from the bubble wall
increases with an increasing difference between the values of $L_{5}$ (or
$\varphi$) between the inside and outside of the bubble\cite{DLM}, so that
photons and massive particle modes with energies $E\ll\delta^{-1}$ inside
either a type I or type II bubble are effectively trapped when $L_{5}$ varies
greatly between these regions. The particle entrapment can help to stabilize
these bubbles from collapsing due to the wall tension\cite{JM1,G-M,JM2}.

Now consider the high temperature limit, with $U$ represented by (\ref{b9}).
We see that at short distances in the $y$ direction the potential $U$ is
positive, but the long distance behavior can be either positive or negative,
depending upon the temperature. More precisely, for $L_{5}\gg\beta$, the
potential $U$ is positive for lower temperatures, $\beta>\beta_{1}$, and
becomes negative for higher temperatures, $\beta<\beta_{1}$.

At very high temperatures $\beta<\beta_{1}$, the positive portion of $U$ at
small $L_{5}\ll\beta_{1}$ must join with the negative portion of $U$ at large
$L_{5}\gg\beta_{1},$ and asymptotically approach zero as $L_{5}\rightarrow
\infty$, indicating the presence of a local minimum somewhere roughly in the
vicinity where the small and large distance parts of $U$ join. We assume the
minimum to be roughly located around $L_{5}\sim O(\beta_{1})$.

On the other hand, at lower temperatures $\beta>\beta_{1}$, the potential has
positive short distance and long distance behaviors, and we infer that $U$ is
a positive monotonically decreasing function of $L_{5}$. Also notice that at
the temperature $\beta=\beta_{1}$ the potential becomes flat at large $L_{5}$
distances. Schematic representations of these basic behaviors are indicated in Fig.2.

Our basic view of the behavior of $U(L_{5},\beta)$ from all this is something
like the following: As the temperature $T$ of the system increases, the
barrier in the low temperature potential shrinks and disappears, and
consequently the state characterizing the system (the expectation value of
$L_{5}$) may tend to roll outward toward larger values of $L_{5}$. However, as
the temperature approaches $T_{1}=1/\beta_{1}$ the potential flattens out and
a lower energy minimum starts to appear at higher temperature, $T>T_{1}$, so
that the state of the system rolls back inward toward this minimum, eventually
settling into this very high temperature vacuum state.

\section{Black Hole Bubbles}

\subsection{Modulus Black Hole Bubble}

For a black hole freely evaporating into empty space, the black hole
temperature is%
\begin{equation}
T_{h}=\frac{1}{8\pi GM_{h}}=\frac{1}{4\pi R_{S}}\label{e16}%
\end{equation}
where $R_{S}=2GM_{h}$ is the Schwarzschild radius. The mass $M_{h}$ of the
black hole decreases at a rate $\dot{M}_{h}\propto-1/(GM_{h})^{2}$, and
consequently, primordial black holes (PBHs) with masses $M_{h}\lesssim
10^{12}kg$ would have evaporated away by now and a PBH with mass $\sim
10^{12}kg$ and size of $\sim1\ fm$ would be quite hot. Near such a PBH
$\beta=1/T$ is small and far away from the PBH $\beta$ is large. The effective
potential $U(L_{5},\beta)$, or $U(\varphi,T)$, would then vary with distance
$r$ from the hole, interpolating between the high temperature and low
temperature forms described above. One then expects the VEV of $L_{5}$, $B$,
and $\varphi$ to vary with $r$. From eq.(\ref{e6}) we have%
\begin{equation}
\square\varphi+\frac{\partial U(\varphi,T)}{\partial\varphi}-\frac
{\partial\mathcal{L}_{4}}{\partial\varphi}=0\label{e17}%
\end{equation}

We can think of this variation in $\varphi(r)$ away from the PBH in terms of a
black hole bubble\cite{Moss,GSS} that surrounds the PBH as the scalar field
passes through a range of values, as would the scalar field of a domain wall
connecting two different vacuum states. The energy density of the bubble wall
depends upon the kinetic and potential contributions from $\varphi$ and the
thickness of the wall depends upon how rapidly the energy density varies. A
thin walled bubble would have a scalar field $\varphi$ changing rapidly over a
small distance, whereas a thick walled bubble would have a more slowly varying
field changing over a larger distance. Let us refer to this type of black hole
bubble, formed by the scalar modulus $\varphi$, simply as a \textquotedblleft
modulus bubble\textquotedblright\ to distinguish it from a \textquotedblleft
symmetry breaking\textquotedblright\ (SB) black hole bubble of the type
originally described by Moss\cite{Moss} that arises from the symmetry
restoration near the PBH for a nonmodulus scalar field.

The forms of the effective potential given by (\ref{b8}) and (\ref{b9}) imply
the existence of a modulus bubble when the interior temperature exceeds
$T_{1}$, i.e., $\beta<\beta_{1}$, provided that%
\begin{equation}%
\begin{array}
[c]{ll}%
1) & T_{1}<T_{h}=\frac{1}{8\pi GM_{h}}\text{, i.e. }\beta_{1}>\kappa^{2}%
M_{h}=4\pi R_{S}\ \ \ \ (\Lambda>0)\\
2) & N_{f}-N_{b}>0\text{ at high }T\text{, and}\\
3) & n_{f}>n_{b}\text{ and }(N_{f}+n_{f})>(N_{b}+n_{b})\text{ at low }T
\end{array}
\label{e17a}%
\end{equation}
(Conditions 2) and 3) seem natural, with the number of fermionic modes
exceeding the number of bosonic ones.) Condition 1) in (\ref{e17a}) imposes a
constraint upon the parameters $\Lambda/\kappa^{2}$ and $2\pi R_{5}$ by
(\ref{b10}), namely,%
\begin{equation}
\frac{\Lambda/\kappa^{2}}{(N_{b}+\frac{15}{16}N_{f})(2\pi R_{5})\frac
{3\zeta(5)}{4\pi^{2}}}<\left(  \frac{1}{\kappa^{2}M_{h}}\right)  ^{5}=\left(
\frac{1}{4\pi R_{S}}\right)  ^{5} \label{e17b}%
\end{equation}
or%
\begin{equation}
2\pi R_{5}>\frac{(\Lambda/\kappa^{2})\left(  4\pi R_{S}\right)  ^{5}}%
{(N_{b}+\frac{15}{16}N_{f})\frac{3\zeta(5)}{4\pi^{2}}} \label{e17c}%
\end{equation}
When these conditions in (\ref{e17a}) hold, the minimum of $U$ near the hole
is negative, with $\varphi=\varphi_{1}$, say, and at asymptotic distances from
the hole ($r\rightarrow\infty,$ $T\rightarrow0$) $U$ is nonnegative in
(\ref{b8}) with $\varphi\rightarrow\varphi_{2}$ asymptotically. Then, whether
a local minimum of the low temperature effective potential exists or not,
$\varphi$ must interpolate between the values $\varphi_{1}$ and $\varphi_{2}$
with an associated variation in $U$ and a nonzero kinetic contribution from
$\varphi$. The kinetic and potential terms contribute to the energy-momentum
tensor $T_{\mu\nu}^{(\varphi)}$. The tensor $T_{\mu\nu}^{(\varphi)}%
\rightarrow0$ as $r\rightarrow\infty$, so that the energy density of the
modulus field is concentrated around the hole. The exact structure of the
bubble, as well as its possible temporal evolution, are dictated by the
solution to (\ref{e17}). Without a solution for $\varphi(r,t)$, the bubble
wall characteristics, such as wall thickness $\delta$, are undetermined, but
will depend upon the length parameters $\beta_{1}$ and $2\pi R_{5}$. By
choosing the asymptotic scale factor $B(\infty)=B_{0}=1$, then the asymptotic
size of the extra dimension is $L_{5}(\infty)=2\pi R_{5}$, which must be large
enough to satisfy the constraint of (\ref{e17c}) for a PBH with a given mass
$M_{h}$ and size $R_{S}$.

We can make a rough estimate of the right hand side of (\ref{e17c}) for a PBH
with $R_{S}\sim1fm\sim.2GeV^{-1}$ using a current value of the cosmological
constant for which $\Lambda/\kappa^{2}\sim10^{-47}GeV^{4}$. Using
$(N_{b}+\frac{15}{16}N_{f})\sim100$, eq.(\ref{e17c}) gives $2\pi
R_{S}>10^{-41}GeV^{-1}$, but this is automatically satisfied for these
parameter choices if we require that $2\pi R_{5}$ be larger than the Planck
length $l_{P}\sim10^{-19}GeV^{-1}$.

A rapidly varying $\varphi$ could have a pronounced effect upon both the
massive and massless particle modes of the radiation emitted by the
evaporating black hole. The particle mass of a (Kaluza-Klein zero mode) boson
or fermion in the effective 4d theory is given\cite{JM1,JM2} by $m=B^{-1/2}%
(\varphi)M$, where $M$ is the mass parameter in the original 5d theory.
Therefore, in a region where $\varphi$ and $B$ decrease with $r$, the mass $m$
increases with $r$, and there is a radially inward attractive force $\vec
{F}\sim-\nabla m$ tending to decelerate a particle moving radially outward.
The particle can become trapped in a region of higher $\varphi$ if it has
insufficient energy to escape. One has the reverse situation if $\varphi$
increases with $r$, i.e. a particle would tend to be accelerated outward,
again toward the region with larger $\varphi$ and smaller $m$. This result
also holds for the Kaluza-Klein excitation modes (see \cite{JM2}). Whatever
the classical particle motion, the probability that the particle escapes
through the wall depends upon the magnitude of the variation in $B$ across the
wall \cite{DLM}. Even if a particle is accelerated radially outward, it may be
reflected back by the wall, with the probability of reflection being given by
the reflection coefficient $\mathcal{R}$. (The coefficient $\mathcal{R}$ is
independent of which side is the incident side.)

The effect of a varying $\varphi$ upon photon propagation was investigated in
\cite{G-M} and \cite{DLM}. An electromagnetic contribution to the effective 4d
theory of the form $-\frac{1}{4}\varepsilon(\varphi)F^{\mu\nu}F_{\mu\nu}$ can
be treated with a dielectric approach where a dielectric function, or
permittivity, in a region of space is defined by $\varepsilon(\varphi
)=B(\varphi)/B_{0}$, with $B_{0}$ being a constant (perhaps asymptotic) value
of the scale factor $B$. The permeability is $\mu=1/\varepsilon$ so that the
index of refraction in a region of space is $n=\sqrt{\varepsilon\mu}=1$ and
the \textquotedblleft impedance\textquotedblright\ is $Z=\sqrt{\mu
/\varepsilon}=1/\varepsilon\propto B^{-1}(\varphi)$. At a sharp boundary
between two different constant values of $B$ (thin wall approximation) the
reflection coefficient is given by\cite{G-M,DLM}%
\begin{equation}
\mathcal{R=}\left(  \frac{\varepsilon_{2}-\varepsilon_{1}}{\varepsilon
_{1}+\varepsilon_{2}}\right)  ^{2}=\left(  \frac{B_{2}-B_{1}}{B_{1}+B_{2}%
}\right)  ^{2}\label{e18}%
\end{equation}
where $\varepsilon_{1,2}$ are the permittivity on the two different sides of
the wall. This result holds for all angles of incidence and for light incident
upon either side of the boundary. When the value of $B(\varphi)$ changes
drastically across the boundary ($B_{2}\ll B_{1}$ or $B_{2}\gg B_{1}$) the
magnitude of the reflection coefficient approaches unity, $\left\vert
\mathcal{R}\right\vert \sim1$. So for a thin-walled bubble where $B$ varies
drastically across the wall from one approximately constant value to another,
the bubble wall is essentially opaque to photons. Photons inside the bubble
would be trapped inside, and would only slowly leak out. The photon pressure
would be exerted radially outward acting to counterbalance the inward pressure
due to the bubble wall tension. For a thick bubble wall, where the photon
wavelength is small compared to wall thickness, $\lambda\ll\delta$, the
results obtained for ordinary (nonmodulus) domain
walls\cite{VSbook,Everett,EDW1,EDW2} lead us to expect that $\mathcal{R}$
becomes small, with $\mathcal{R}\rightarrow0$ as $\lambda\rightarrow0$. This
was indeed found to be the case in ref.\cite{DLM}. We therefore expect an
entrapment of \textquotedblleft low energy\textquotedblright\ ($\omega
\lesssim\delta^{-1})$ photons inside the bubble, while \textquotedblleft high
energy\textquotedblright\ ($\omega\gtrsim\delta^{-1}$) photons escape. The
same qualitative statements hold for the case of massive particle
modes\cite{DLM}, and in the case that one side of the wall becomes
kinematically inaccessible to sufficiently low energy particles (say,
$E<m_{1}$ or $E<m_{2}$), there is a total reflection from the wall.

We form the following rough picture for a modulus black hole bubble. Near the
horizon of a very hot PBH ($T\gtrsim T_{1}$) we envision the vev associated
with $L_{5}$ to be on the order of $\beta_{1}=1/T_{1}$ with an associated
scale factor $B_{1}=L_{5}/(2\pi R_{5})\sim O(\beta_{1})/(2\pi R_{5})$.
However, asymptotically $T$ approaches a value $L_{5}(\infty)=2\pi R_{5}$ for
$B(\infty)=B_{0}=1$. The ratio $B_{hor}/B_{0}\sim$ $B_{1}/B_{0}\sim
O(\beta_{1})/(2\pi R_{5})\gtrsim R_{S}/R_{5}$ depends on the value of
$\beta_{1}$. With a large change in $B(\varphi)$ we have a black hole bubble
forming around the PBH which tends to trap lower energy photons and massive
particles inside, due to a large reflection coefficient $\mathcal{R}$, which
by (\ref{e18}) can be near unity. Massive particles that are kinematically
forbidden to escape are totally reflected. However, for higher energy modes
with energies $E\gg\delta^{-1}$, the reflectivity becomes small and particles
escape. The decrease in $\mathcal{R}$ becomes pronounced at an energy
$E\sim\delta^{-1}$ \cite{DLM}. The structure and evolution of the bubble
require information about the bubble unavailable to us, and can only be
described qualitatively, at best. For instance, it would be of interest to
know the rate of photon and particle reabsorption by the black hole and the
back reaction effects on the PBH evaporation. Also, as pointed out by
Moss\cite{Moss}, there seems to be an intermediate case here between a black
hole radiating freely into empty space and a black hole in thermal
equilibrium. At any rate, there appears to be a strong possibility that the
low energy ($E\ll\delta^{-1}$) photon and particle components of the PBH
thermal spectrum will not be observed asymptotically, whereas the high energy
particles ($E\gg\delta^{-1}$) will penetrate the modulus bubble wall. The
presence of a modulus black hole bubble will therefore alter the emission
spectrum seen by an outside observer.

The possibity might be entertained that a PBH that has evaporated away leaves
behind a modulus bubble filled with various types of particles. The tendency
for the bubble to shrink is countered by the particle pressure exerted outward
on the bubble wall. The characteristics and evolution of a thin walled bubble
with an interior temperatutre $T$, with a particle energy density dominated by
the effectively massless radiation modes, is expected to resemble those
previously described for radiation filled dimension bubbles\cite{JM2}. There
the bubble mass $\mathcal{M}$ depends on the particle radiation energy density
$\rho_{rad}=AT^{4}$, the $\varphi$-dependent interior energy density $\lambda
$, the wall tension $\sigma$, and the bubble radius $R_{B}$. The minimization
of $\mathcal{M}$ determines the equilibrium bubble size and mass. In
\cite{JM2} these were found for a stable bubble  to be given by%
\begin{equation}%
\begin{array}
[c]{ll}%
R_{B}=\dfrac{6\sigma}{(\rho_{rad}-3\lambda)},\ \ \ \ \  & \mathcal{M}%
=12\pi\sigma R_{B}^{2}\left(  \dfrac{\rho_{rad}-\frac{1}{3}\lambda}{\rho
_{rad}-3\lambda}\right)
\end{array}
\label{bub1}%
\end{equation}
provided that $\rho_{rad}/3>\lambda$.

\subsection{Symmetry Breaking Black Hole Bubble}

The SB black hole bubble proposed by Moss\cite{Moss} arises from a high
temperature region near the PBH entering a symmetric phase with a Higgs field
vev $\phi=0$ bounded by a broken symmetric phase further away where $\phi
\neq0$. The bubble is bounded by a domain wall that separates these two
phases. We then have the possibility that, due to the Higgs mechanism,
particles can be massless in the high $T$ symmetric phase inside the bubble,
but have nonzero rest mass $m$ outside the bubble. Lower energy particles
emitted by the PBH with $E<m$ would become trapped inside and tend to resist
the shrinkage of the bubble. More generally, particles incident upon the
bubble wall typically suffer some degree of reflection from the wall. The
reflection probability is expected to be larger for lower energy particles
(wavelength $\gg$ wall thickness) and smaller for high energy particles
(wavelength $\ll$ wall thickness). (See, e.g., refs.\cite{Everett,EDW1,EDW2}.)

Photons are not trapped by the same dynamical mechanism, however, as they are
massless in both phases. However, photons can be reflected from a bubble
domain wall that is built from an isodoublet scalar field, as in electroweak
theory, where one scalar component picks up a nonzero vev in the broken
symmetric phase. When the mixing between the vector fields constituting the
physical photon field changes across the domain wall, then photons are also
found to suffer some degree of reflection from the wall. Again, relatively low
energy photons (wavelength $\gg$ wall thickness) are strongly reflected and
relatively high energy photons (wavelength $\ll$ wall thickness) are strongly
transmitted\cite{Everett,EDW1,EDW2}.

Therefore the bubble wall prevents the immediate escape of some of the
normally massive particles, and possibly photons (for domain walls built from
nonisoscalar scalar fields). A pressure is exerted on the wall, tending to
halt its collapse. Assuming the dynamical time scales associated with bubble
equilibration to be short compared to the PBH evaporation time, the bubble
reaches an equilibrium state where the forces acting on the wall are balanced.
Moss conjectured that such a bubble would exhibit a $\gamma$-ray luminosity
which could be enhanced at certain energies.

\subsection{Double Bubble}

If particle theories based upon spontaneous symmetry breaking are correct,
\textit{and} if there exists one or more extra space dimensions that can be
inhomogeneously compactified, then it seems reasonable to speculate that (for
certain parameter ranges) \textit{both} types of black hole bubbles may
enclose a hot PBH. In other words, we envision a situation wherein at least a
portion of the interior of one type of bubble coincides with the interior of
the other type of bubble as well. The PBH is enclosed by a double bubble. In
the inner portion of the double bubble the rest mass of a normally (low
temperature) massive particle vanishes, so that low energy particles get
trapped. But photons can also get trapped by a photon-opaque thin walled
modulus bubble with a large variation in the scale factor $B$ across the wall.
The double bubble can therefore provide a substantial pressure serving to
stabilize the bubble against immediate collapse. The amount and
distinctiveness of the radiation emerging from the black hole and bubble will
then depend upon the characteristics of the bubble walls, black hole
temperature and evaporation rate, and approach to thermal equilibrium. The end
result could be a nonnegligible deviation from the predictions for a black
hole freely radiating into empty space. Furthermore, if black hole bubbles are
endpoints of the Hawking radiation, smaller mass PBHs that have already
evaporated may have left behind metastable bubbles that have not yet decayed,
which could contribute to the dark matter of the universe.

\section{Summary}

Primordial black holes of mass $M_{h}\gtrsim10^{12}$ kg may still be present
in the universe today, and some may be quite hot, depending upon the mass. The
temperature of the plasma around a black hole decreases with distance away
from it, so that for sufficiently high temperature PBHs there can be a region
of restored symmetry (e.g., electroweak symmetry) near the hole and a region
of broken symmetry further away. These regions are separated by a domain wall
which bounds a symmetry breaking (SB) black hole bubble. The bubble wall is
expected to possess a reflection coefficient which decreases with increasing
particle energy, but for particles with subcritical energies the reflection
from the wall back into the bubble is total. So some of the Hawking radiation
is blocked from escaping the black hole bubble.

If there is an extra space dimension which is compactified, the
compactification can become inhomogeneous near a hot PBH. For the model
considered here with one extra space dimension compactified on a circle, the
inhomogeneous compactification gives rise to a second type of black hole
bubble (modulus black hole bubble) characterized by a scale factor
$B=\sqrt{-\tilde{g}_{55}}$ which varies with distance, provided that the
conditions listed in (\ref{e17a}) are satisfied, i.e., for the portion of
parameter space where $\Lambda/\kappa^{2}$ and $2\pi R_{5}$ are constrained by
(\ref{e17c}). When the values of $B$ inside and outside the bubble wall vary
greatly, the reflection coefficient $\mathcal{R}(E)$ can approach unity for
particles and photons with energies $E\ll\delta^{-1}$, where the bubble wall
width $\delta$ depends upon model parameters.

Both types of black hole bubble may be produced, so that the PBH is enclosed
within a black hole double bubble. The double bubble walls can impede the
passage of both massive and massless particles produced by the PBH from
within, so that a pressure tends to build up within the bubble. The attendant
description is necessarily qualitative and somewhat speculative, since the
actual characteristics of the bubble walls, back reaction on the black hole
through particle reabsorption, and approach to thermal equilibrium, etc. are
model-dependent and unknown. However, we conclude that if particle theories
based upon spontaneous symmetry breaking are correct and/or there exists one
or more extra space dimension(s) that becomes inhomogeneously compactified
near a hot PBH, then the spectrum of radiation coming from the black hole
could be significantly disturbed from what is expected for a blackbody. The
presence of black hole bubbles due to symmetry restoration and/or extra
dimensions may effectively disguise or hide evaporating black holes. PBHs
created with masses $M_{h}\lesssim10^{12}$ kg that have already evaporated
away may have left behind metastable black hole bubbles which, if still in
existence, contribute to the dark matter of the universe.

\bigskip

\textbf{Ackowledgement:} I thank Eduardo Guendelman for comments.

\appendix{}

\section{The 4d Effective Potential}

The 4d effective potential $U(L_{5},\beta)$ \ has a contribution from a
cosmological constant $\Lambda$ in the 5d theory, along with a contribution
from the Rubin-Roth (RR) potential $\tilde{\Gamma}(L_{5},\beta)$ for one-loop
quantum corrections at finite temperature due to Casimir effects for bosons
and fermions\cite{RR1,RR2}. Actually $\tilde{\Gamma}(L_{5},\beta)$ is a
Euclidean action density for a 5d theory with a flat 4d Jordan frame (JF)
spacetime. This can be translated to an action density $\tilde{S}(L_{5}%
,\beta)$ for a flat 4d Einstein frame (EF). The 4d RR potential $U_{RR}%
(L_{5},\beta)$ can be extracted from $\tilde{\Gamma}(L_{5},\beta)$ or
$\tilde{S}(L_{5},\beta)$ and the full 4d EF effective potential $U(L_{5}%
,\beta)$ is built from the $U_{RR}$ and $\Lambda$ pieces.

\subsection{Rubin-Roth Effective Actions}

\textit{Jordan and Einstein frames: }The 5d spacetime is described by%
\begin{equation}
ds^{2}=\tilde{g}_{\mu\nu}dx^{\mu}dx^{\nu}-B^{2}dy^{2}=B^{-1}g_{\mu\nu}dx^{\mu
}dx^{\nu}-B^{2}dy^{2}%
\end{equation}
The 4d JF metric is $\tilde{g}_{\mu\nu}$ (the $\mu\nu$ part of $\tilde{g}%
_{MN}$), and the 4d EF metric is $g_{\mu\nu}$,%
\begin{equation}
\tilde{g}_{\mu\nu}=B^{-1}g_{\mu\nu},\ \ \ \ \ \tilde{g}_{55}=-B^{2}\label{ap2}%
\end{equation}
and $\tilde{g}_{5}=\det(\tilde{g}_{MN})=(-\tilde{g}_{4})(-\tilde{g}%
_{55})=B^{2}\left(  -\tilde{g}_{4}\right)  $. So%
\begin{equation}
\sqrt{\tilde{g}_{5}}=B\sqrt{-\tilde{g}_{4}}=B^{-1}\sqrt{-g_{4}}\label{ap3}%
\end{equation}
There are two different spacetimes being considered here, one with a flat 4d
JF and one with a flat 4d EF;%
\begin{align}
\text{flat JF} &  \text{: \ \ \ }\tilde{g}_{\mu\nu}=\eta_{\mu\nu}%
,\ \ \sqrt{-\tilde{g}_{4}}=1,\ \ \sqrt{\tilde{g}_{5}}=B\label{ap4}\\
\text{flat EF} &  \text{:\ \ \ \ }g_{\mu\nu}=\eta_{\mu\nu},\ \ \sqrt{-g_{4}%
}=1,\ \ \sqrt{\tilde{g}_{5}}=B^{-1}\label{ap5}%
\end{align}

We want to relate the Rubin-Roth (RR) actions $\Gamma,\tilde{\Gamma}$
evaluated in a flat JF to the actions $S,\tilde{S}$ evaluated in a flat EF.
$\Gamma$ and $S$ are effective Euclidean actions and $\tilde{\Gamma}$ and
$\tilde{S}$ are action (3-) densities, defined by%
\begin{equation}
\tilde{\Gamma}=\frac{\Gamma}{\int d^{3}x},\ \ \ \ \ \tilde{S}=\frac{S}{\int
d^{3}x}\label{ap6}%
\end{equation}
Denote the 5d Euclidean action measure by  $d^{5}x_{E}$, with a Wick rotation
of the $t$ coordinate.

\textit{RR Effective Actions, Flat JF: }Although we are interested in $x^{\mu
}$-dependent $T$ and $B$, we treat them as constant parameters in the
evaluation of the effective action. For a flat JF we use $\sqrt{\tilde{g}_{5}%
}=B$. Introduce a 5d effective potential $V_{5}$ (see, e.g.,
Appelquist-Chodos\cite{AC2}), and write a 5d effective action%
\begin{equation}
\Gamma=V_{5}\int d^{5}x_{E}\sqrt{\tilde{g}_{5}}=BV_{5}\int_{0}^{\beta}%
d\tau\int_{0}^{2\pi R_{5}}dy\int d^{3}x=\int d^{3}x\tilde{\Gamma
},\ \label{ap7}%
\end{equation}
so that, with $L_{5}\equiv(2\pi R_{5})B$, we have%

\begin{equation}
\tilde{\Gamma}=\beta L_{5}V_{5},\ \ \ \ \ V_{5}=\dfrac{\tilde{\Gamma}}{\beta
L_{5}}\label{ap8}%
\end{equation}
The $\tilde{\Gamma}$ are the effective actions given by Rubin and Roth for
1-loop quantum effects at finite $T$. They are obtained for a flat Jordan
frame spacetime. For the ($N_{B}=5$) graviton degrees of freedom in the zero
temperature ($\beta\rightarrow\infty$) limit, from (\ref{ap8}) and the RR
effective potential $\tilde{\Gamma}$ we recover the Appelquist-Chodos
potential\cite{AC2}, $V_{AC}$, i.e.,
\begin{equation}
V_{5}=\frac{\tilde{\Gamma}}{\beta L_{5}}=\frac{1}{\beta L_{5}}\left(
-\frac{15}{4\pi^{2}}\zeta(5)\frac{\beta}{L_{5}^{4}}\right)  =-\frac{15}%
{4\pi^{2}}\zeta(5)\frac{1}{L_{5}^{5}}=V_{AC}\label{ap8a}%
\end{equation}

\textit{RR Effective Actions, Flat EF: }For a flat EF we use $\sqrt{\tilde
{g}_{5}}=B^{-1}$, $\sqrt{-g_{4}}=1$. Denote the effective action in this space
by $S$, and the density by $\tilde{S}=S/\int d^{3}x$. Then%
\begin{equation}
S=V_{5}\int d^{5}x_{E}\sqrt{\tilde{g}_{5}}=B^{-1}V_{5}\int_{0}^{\beta}%
d\tau\int_{0}^{2\pi R_{5}}dy\int d^{3}x=\frac{\beta V_{5}L_{5}}{B^{2}}\int
d^{3}x\label{ap9}%
\end{equation}
Therefore, from (\ref{ap8})%

\begin{equation}
\tilde{S}=\dfrac{\beta L_{5}}{B^{2}}V_{5}=\dfrac{\tilde{\Gamma}}{B^{2}}=(2\pi
R_{5})^{2}\dfrac{\tilde{\Gamma}}{L_{5}^{2}}\label{ap10}%
\end{equation}
The density $\tilde{S}$ is the corresponding effective RR action density for a
flat Einstein frame spacetime.

\subsection{Form of the 4d Effective Potential $U(L_{5},\beta)$}

The 5d action is%
\begin{equation}%
\begin{array}
[c]{ll}%
S & =\int d^{5}x\sqrt{\tilde{g}_{5}}\left\{  \dfrac{1}{2\kappa_{5}^{2}}\left(
\tilde{R}_{5}-2\Lambda\right)  +\mathcal{L}_{5}\right\}  \\
& =\int d^{4}x\sqrt{\tilde{g}_{4}}B\left\{  \dfrac{1}{2\kappa^{2}}\left(
\tilde{R}_{5}-2\Lambda\right)  +(2\pi R_{5})\mathcal{L}_{5}\right\}
\end{array}
\label{ap11}%
\end{equation}
giving the dimensionally reduced 4d EF effective action%
\begin{equation}
S=\int d^{4}x\sqrt{-g}\left\{  \frac{1}{2\kappa^{2}}R+\frac{3}{4\kappa^{2}%
}\left(  \frac{\nabla B}{B}\right)  ^{2}+B^{-1}\left[  (2\pi R_{5}%
)\mathcal{L}_{5}-\frac{\Lambda}{\kappa^{2}}\right]  \right\}  \label{ap12}%
\end{equation}
The 4d EF effective Lagrangian arising from the $\mathcal{L}_{5}$ and
$\Lambda$ terms is%
\begin{equation}
\mathcal{L}_{4},_{eff}=B^{-1}\left[  (2\pi R_{5})\mathcal{L}_{5}-\frac
{\Lambda}{\kappa^{2}}\right]  \label{ap13}%
\end{equation}
We set $\mathcal{L}_{5}=-V_{5}$, where $V_{5}$ is the 5d effective potential
in (\ref{ap8}) and (\ref{ap10}), and then identify the 4d EF effective
potential $\mathcal{L}_{4},_{eff}=-U$:%
\begin{align}
U &  =B^{-1}\left[  (2\pi R_{5})V_{5}+\frac{\Lambda}{\kappa^{2}}\right]
=U_{RR}+U_{\Lambda},\label{ap14}\\
U_{RR} &  =B^{-1}(2\pi R_{5})V_{5},\ \ \ \ \ U_{\Lambda}=B^{-1}\frac{\Lambda
}{\kappa^{2}}\label{ap15}%
\end{align}
where the RR and $\Lambda$ parts are combined to give the total potential $U$.
The effective potential $U=U_{RR}+U_{\Lambda}$ can be written in terms of $B$
or $L_{5}$:%
\begin{equation}%
\begin{array}
[c]{l}%
U_{RR}=\dfrac{\tilde{\Gamma}}{\beta B^{2}}=(2\pi R_{5})^{2}\dfrac
{\tilde{\Gamma}}{\beta L_{5}^{2}}\\
U_{\Lambda}=\dfrac{\Lambda/\kappa^{2}}{B}=(2\pi R_{5})\dfrac{\Lambda
/\kappa^{2}}{L_{5}}\\
U(L_{5},\beta)=(2\pi R_{5})^{2}\dfrac{\tilde{\Gamma}}{\beta L_{5}^{2}}+(2\pi
R_{5})\dfrac{\Lambda/\kappa^{2}}{L_{5}}%
\end{array}
\label{ap16}%
\end{equation}

\section{Asymptotic Forms of $U(L_{5},\beta)$}

\subsection{Rubin-Roth part}

Rubin and Roth have compiled a table (Table 1 in ref.\cite{RR2}) of effective
potential contributions per degree of freedom for bosons and fermions (for
untwisted and twisted fields). Here, we use only the untwisted contributions.
Tables I and II list these contributions for the ultrarelativistic and
nonrelativistic limits. The effective potential per degree of freedom is
denoted here by $\bar{\Gamma}=\tilde{\Gamma}/($degree freedom) and $M$ refers
to particle mass. (The table entries are numbered for use below.)

\begin{center}%
\begin{tabular}
[c]{|c|rl|}\hline
eff pot $\tilde{\Gamma}$ & Ultrarel & ($M\ll T$)\\\hline
\multicolumn{1}{|l|}{$%
\begin{array}
[c]{l}%
\text{{\small per degree}}\\
\text{{\small of freedom}}%
\end{array}
$} & \multicolumn{1}{|c}{$L_{5}\ll\beta\ll1/M$} & \multicolumn{1}{|l|}{$%
\begin{array}
[c]{l}%
\beta\ll L_{5}\ll1/M\\
\multicolumn{1}{c}{\text{and}}\\
\beta\ll1/M\ll L_{5}%
\end{array}
$}\\\hline\hline
\multicolumn{1}{|l|}{$\bar{\Gamma}_{b}$} & \multicolumn{1}{|l}{$-\frac
{3\zeta(5)}{4\pi^{2}}\frac{\beta}{L_{5}^{4}}$ \fbox{{\small 1}}} &
\multicolumn{1}{|l|}{$-\frac{3\zeta(5)}{4\pi^{2}}\frac{L_{5}}{\beta^{4}}$
\fbox{{\small 7}}}\\\hline
\multicolumn{1}{|l|}{$\bar{\Gamma}_{f}$} & \multicolumn{1}{|l}{$\frac
{3\zeta(5)}{4\pi^{2}}\frac{\beta}{L_{5}^{4}}$ \fbox{{\small 2}}} &
\multicolumn{1}{|l|}{$-\frac{3\zeta(5)}{4\pi^{2}}\left(  \frac{15}{16}\right)
\frac{L_{5}}{\beta^{4}}$ \fbox{{\small 8}}}\\\hline
\end{tabular}

{\small Table I. Effective potential (action density) }$\tilde{\Gamma}$
{\small per degree of freedom for ultrarelativistic modes.}

\bigskip%

\begin{tabular}
[c]{|c|lll|}\hline
eff pot $\tilde{\Gamma}$ &  & Nonrel ($M\gg T$) & \\\hline
\multicolumn{1}{|l|}{$%
\begin{array}
[c]{l}%
\text{{\small per degree}}\\
\text{{\small of freedom}}%
\end{array}
$} & $L_{5}\ll1/M\ll\beta$ & \multicolumn{1}{|l}{$1/M\ll L_{5}\ll\beta$} &
\multicolumn{1}{|l|}{$1/M\ll\beta\ll L_{5}$}\\\hline\hline
\multicolumn{1}{|l|}{$\bar{\Gamma}_{b}$} & $-\frac{3\zeta(5)}{4\pi^{2}}%
\frac{\beta}{L_{5}^{4}}$ \fbox{{\small 3}} & \multicolumn{1}{|l}{$-\frac
{M^{2}}{4\pi^{2}}\frac{\beta}{L_{5}^{2}}e^{-ML_{5}}$ \fbox{{\small 5}}} &
\multicolumn{1}{|l|}{$-\frac{M^{2}}{4\pi^{2}}\frac{\beta}{L_{5}^{2}}%
e^{-M\beta}$ \fbox{{\small 9}}}\\\hline
\multicolumn{1}{|l|}{$\bar{\Gamma}_{f}$} & $\frac{3\zeta(5)}{4\pi^{2}}%
\frac{\beta}{L_{5}^{4}}$ \fbox{{\small 4}} & \multicolumn{1}{|l}{$\frac{M^{2}%
}{4\pi^{2}}\frac{\beta}{L_{5}^{2}}e^{-ML_{5}}$ \fbox{{\small 6}}} &
\multicolumn{1}{|l|}{$-\frac{M^{2}}{4\pi^{2}}\frac{\beta}{L_{5}^{2}}%
e^{-M\beta}$ \fbox{{\small 10}}}\\\hline
\end{tabular}

{\small Table II. Effective potential (action density) }$\tilde{\Gamma}$
{\small per degree of freedom for nonrelativistic modes.}

\end{center}

The following notation is used to label particle modes:
\[%
\begin{array}
[c]{lll}%
N_{b(f)}= & \text{number of \textit{effectively massless} modes at temperature
}T\text{;} & M\ll T,\ \beta\ll1/M\\
n_{b(f)}= & \text{number of \textit{massive} modes at temperature }T\text{;} &
M\gg T,\ \beta\gg1/M
\end{array}
\]
E.g., at low $T$ the effectively massless modes may be the exactly massless
modes (like graviton, photon), and very light fermions and scalars (like $\nu
$'s); very massive modes would become relatively suppressed through factors
like $e^{-ML_{5}}$ and $e^{-M\beta}$. The various terms in tables I and II are
represented by $\bar{\Gamma}_{i}$ for term \fbox{i}.

\subsection{Low\textit{ }$T$ Limit ($T\rightarrow0$)}

We consider $T\approx0$, $\beta\rightarrow\infty$, but finite. For effectively
\textit{massless modes}, $M\ll T$, $\beta\ll\frac{1}{M}$,%
\[
\bar{\Gamma}\sim\left\{
\begin{array}
[c]{ccc}%
\text{terms 1,2} & \left(  \bar{\Gamma}_{1},\bar{\Gamma}_{2}\right)  , &
L_{5}\ll\beta\ll1/M\\
\text{terms 7,8} & \left(  \bar{\Gamma}_{7},\bar{\Gamma}_{8}\right)  , &
\beta\ll L_{5}\ll1/M
\end{array}
\right\}
\]%
\begin{equation}
\tilde{\Gamma}\approx\left\{
\begin{array}
[c]{cc}%
(N_{f}-N_{b})\frac{3\zeta(5)}{4\pi^{2}}\frac{\beta}{L_{5}^{4}}, & L_{5}%
\ll\beta\ll1/M\\
-(N_{b}+\frac{15}{16}N_{f})\frac{3\zeta(5)}{4\pi^{2}}\frac{L_{5}}{\beta^{4}%
}, & \beta\ll L_{5}\ll1/M
\end{array}
\right\}  ,\ \ \ \ \ M\ll T,\ \beta\ll\frac{1}{M}\label{ap17}%
\end{equation}

For \textit{massive modes}, $M\gg T$, $\beta\gg\frac{1}{M}$, ($\beta
\rightarrow\infty$, but finite),%
\[
\bar{\Gamma}\sim\left\{
\begin{array}
[c]{ccc}%
\text{terms 3,4} & \left(  \bar{\Gamma}_{3},\bar{\Gamma}_{4}\right)  , &
L_{5}\ll\frac{1}{M}\ll\beta\\
\text{terms 5,6} & \left(  \bar{\Gamma}_{5},\bar{\Gamma}_{6}\right)  , &
\frac{1}{M}\ll L_{5}\ll\beta
\end{array}
\right\}
\]%
\begin{equation}
\tilde{\Gamma}\approx\left\{
\begin{array}
[c]{cc}%
(n_{f}-n_{b})\frac{3\zeta(5)}{4\pi^{2}}\frac{\beta}{L_{5}^{4}}, & L_{5}%
\ll\frac{1}{M}\ll\beta\\
(n_{f}-n_{b})\frac{M^{2}}{4\pi^{2}}\frac{\beta}{L_{5}^{2}}e^{-ML_{5}}, &
\frac{1}{M}\ll L_{5}\ll\beta
\end{array}
\right\}  ,\ \ \ \ M\gg T\text{,\ }\beta\gg\frac{1}{M} \label{ap18}%
\end{equation}

\textit{Total }$\tilde{\Gamma}$\textit{ for all modes at low }$T$: add all of
the $L_{5}\ll\frac{1}{M}$ terms (but drop the bottom terms $\bar{\Gamma}_{7}$,
$\bar{\Gamma}_{8}$ in (\ref{ap17}), since these terms $\rightarrow0$ as
$\beta\rightarrow\infty$) and add all of the $L_{5}\gg\frac{1}{M}$ terms to
get a total $\tilde{\Gamma}$ for small and large $L_{5}$ at low $T$:%
\begin{equation}
\tilde{\Gamma}\approx\left\{
\begin{array}
[c]{ll}%
\lbrack(N_{f}+n_{f})-(N_{b}+n_{b})]\frac{3\zeta(5)}{4\pi^{2}}\frac{\beta
}{L_{5}^{4}}, & L_{5}\ll1/M\\
(n_{f}-n_{b})\frac{M^{2}}{4\pi^{2}}\frac{\beta}{L_{5}^{2}}e^{-ML_{5}}, &
L_{5}\gg1/M
\end{array}
\right\}  \ \ \ \ \ \text{(}\tilde{\Gamma}\ \text{at low }T\text{)}
\label{ap19}%
\end{equation}

\subsection{High $T$ Limit\ \ \ \ ($T\gg M$)}

For the high $T$ limit ($T\gg M,$ $\beta\ll\frac{1}{M}$) we consider only the
ultrarelativistic effectively massless modes, as they dominate the massive
ones for the asymptotic forms.%

\[
\bar{\Gamma}\sim\left\{
\begin{array}
[c]{ll}%
\bar{\Gamma}_{1},\bar{\Gamma}_{2}, & \text{(small }L_{5}\text{)\ \ \ }L_{5}%
\ll\beta\ll\frac{1}{M}\\
\bar{\Gamma}_{7},\bar{\Gamma}_{8}, & \text{(large }L_{5}\text{)\ \ }L_{5}%
\gg\beta
\end{array}
\right\}
\]

\begin{equation}
\tilde{\Gamma}\approx\left\{
\begin{array}
[c]{ll}%
(N_{f}-N_{b})\frac{3\zeta(5)}{4\pi^{2}}\frac{\beta}{L_{5}^{4}}, & \text{(small
}L_{5}\text{)\ \ \ }L_{5}\ll\beta\\
-(N_{b}+\frac{15}{16}N_{f})\frac{3\zeta(5)}{4\pi^{2}}\frac{L_{5}}{\beta^{4}%
}, & \text{(large }L_{5}\text{)\ \ }L_{5}\gg\beta
\end{array}
\right\}  \text{\ \ \ (}\Gamma_{RR}\ \ \text{at high }T\text{, }T\gg M\text{)}
\label{ap20}%
\end{equation}

\subsection{Total Effective Potential}

From (\ref{ap16}), (\ref{ap19}) and (\ref{ap20}), the asymptotic forms of the
total effective potential $U$ can be written out for the low and high
temperature limits. For low $T$,%

\begin{equation}
U\approx\left\{
\begin{array}
[c]{ll}%
\lbrack(N_{f}+n_{f})-(N_{b}+n_{b})](2\pi R_{5})^{2}\frac{3\zeta(5)}{4\pi^{2}%
}\frac{1}{L_{5}^{6}}+\frac{(2\pi R_{5})\Lambda}{\kappa^{2}L_{5}}, & L_{5}%
\ll1/M\\
(n_{f}-n_{b})(2\pi R_{5})^{2}\frac{M^{2}}{4\pi^{2}}e^{-ML_{5}}\frac{1}%
{L_{5}^{4}}+\frac{(2\pi R_{5})\Lambda}{\kappa^{2}L_{5}}, & L_{5}\gg1/M
\end{array}
\right\}  \text{ \ \ \ (Low }T\text{)}\label{ap21}%
\end{equation}
where $N=\#$ massless modes, $n=\#$ of massive modes, each with mass $\sim M$.
This $U$ is $\beta$ independent, $\beta\gg\frac{1}{M}$, with $T\rightarrow
0^{+},\ \beta\rightarrow\infty$, but finite. For high $T$,%

\begin{equation}
U\approx\left\{
\begin{array}
[c]{ll}%
\lbrack(N_{f}-N_{b})](2\pi R_{5})^{2}\frac{3\zeta(5)}{4\pi^{2}}\frac{1}%
{L_{5}^{6}}+\frac{(2\pi R_{5})\Lambda}{\kappa^{2}L_{5}}, & L_{5}\ll\beta\\
-(N_{b}+\frac{15}{16}N_{f})(2\pi R_{5})^{2}\frac{3\zeta(5)}{4\pi^{2}}\frac
{1}{\beta^{5}L_{5}}+\frac{(2\pi R_{5})\Lambda}{\kappa^{2}L_{5}}, & L_{5}%
\gg\beta
\end{array}
\right\}  \text{ \ \ \ \ (High }T\text{)}\label{ap22}%
\end{equation}
with $T\gg M,\ \ \beta\ll\frac{1}{M}$. (relativistic modes) \ We assume
$N_{f}^{Tot}>N_{b}^{Tot}$ for all cases.

\ \ \ We can define the temperature parameters $T_{1}$ and $\beta_{1}$,%
\begin{equation}
T_{1}=\frac{1}{\beta_{1}}=\left\{  \frac{\Lambda/\kappa^{2}}{(N_{b}+\frac
{15}{16}N_{f})(2\pi R_{5})\frac{3\zeta(5)}{4\pi^{2}}}\right\}  ^{1/5}
\label{ap23}%
\end{equation}
which are real, provided that $\Lambda>0$. There are three parameters, $(2\pi
R_{5}),\Lambda/\kappa^{2},$ and $\beta_{1}$, related by the definition of
$\beta_{1}$, leaving two independent parameters, e.g., $(2\pi R_{5})$ and
$\beta_{1}$. The high $T$ effective potential, in terms of $\beta_{1}$, is
then%
\begin{equation}
U\approx\left[  (N_{b}+\frac{15}{16}N_{f})(2\pi R_{5})^{2}\frac{3\zeta
(5)}{4\pi^{2}}\right]  \frac{1}{\beta_{1}^{5}}\left\{
\begin{array}
[c]{cc}%
\frac{(N_{f}-N_{b})}{(N_{b}+\frac{15}{16}N_{f})}\frac{\beta_{1}^{5}}{L_{5}%
^{6}}+\frac{1}{L_{5}}, & L_{5}\ll\beta\\
\left(  1-\frac{\beta_{1}^{5}}{\beta^{5}}\right)  \frac{1}{L_{5}}, & L_{5}%
\gg\beta
\end{array}
\right\}  \text{\ \ \ \ (high }T\text{)} \label{ap24}%
\end{equation}

\subsection{Summary}

The limiting forms for the temperature-dependent 4d effective potential
$U(L_{5},\beta)$ are given by%

\begin{equation}
U\approx\left\{
\begin{array}
[c]{ll}%
\lbrack(N_{f}+n_{f})-(N_{b}+n_{b})](2\pi R_{5})^{2}\frac{3\zeta(5)}{4\pi^{2}%
}\frac{1}{L_{5}^{6}}+\frac{(2\pi R_{5})\Lambda}{\kappa^{2}L_{5}}, & L_{5}%
\ll1/M\\
(n_{f}-n_{b})(2\pi R_{5})^{2}\frac{M^{2}}{4\pi^{2}}e^{-ML}\frac{1}{L_{5}^{4}%
}+\frac{(2\pi R_{5})\Lambda}{\kappa^{2}L_{5}}, & L_{5}\gg1/M
\end{array}
\right\}  \text{\ \ \ (low }T\text{)} \label{ap25}%
\end{equation}

\begin{equation}
U\approx\left[  (N_{b}+\frac{15}{16}N_{f})(2\pi R_{5})^{2}\frac{3\zeta
(5)}{4\pi^{2}}\right]  \frac{1}{\beta_{1}^{5}}\left\{
\begin{array}
[c]{cc}%
\frac{(N_{f}-N_{b})}{(N_{b}+\frac{15}{16}N_{f})}\frac{\beta_{1}^{5}}{L_{5}%
^{6}}+\frac{1}{L_{5}}, & L_{5}\ll\beta\\
\left(  1-\frac{\beta_{1}^{5}}{\beta^{5}}\right)  \frac{1}{L_{5}}, & L_{5}%
\gg\beta
\end{array}
\right\}  \ \ \ \ (\text{high }T) \label{ap26}%
\end{equation}

\newpage

\begin{center}
\textbf{Figure Captions}
\end{center}

\bigskip

\textbf{Figure 1:} Schematic depiction of $U$ vs $L_{5}$ for low temperature

\bigskip

\textbf{Figure 2:} Schematic depictions of $U$ vs $L_{5}$ for high temperature
$T<T_{1}$ (upper curve) and for very high temperature $T>T_{1}$ (lower curve)

\bigskip%
%TCIMACRO{\FRAME{dtbpFU}{3.5241in}{2.0686in}{0pt}{\Qcb{Figure 1}}{}%
%{fig1.eps}{\special{ language "Scientific Word";  type "GRAPHIC";
%width 3.5241in;  height 2.0686in;  depth 0pt;  original-width 3.4973in;
%original-height 8.0652in;  cropleft "0";  croptop "1";  cropright "1";
%cropbottom "0.7469";  filename '../Sb7/fig1.eps';file-properties "XNPEU";}}}%
%BeginExpansion
\begin{center}
\includegraphics[
trim=0.000000in 6.023898in 0.000000in 0.000000in,
height=2.0686in,
width=3.5241in
]%
{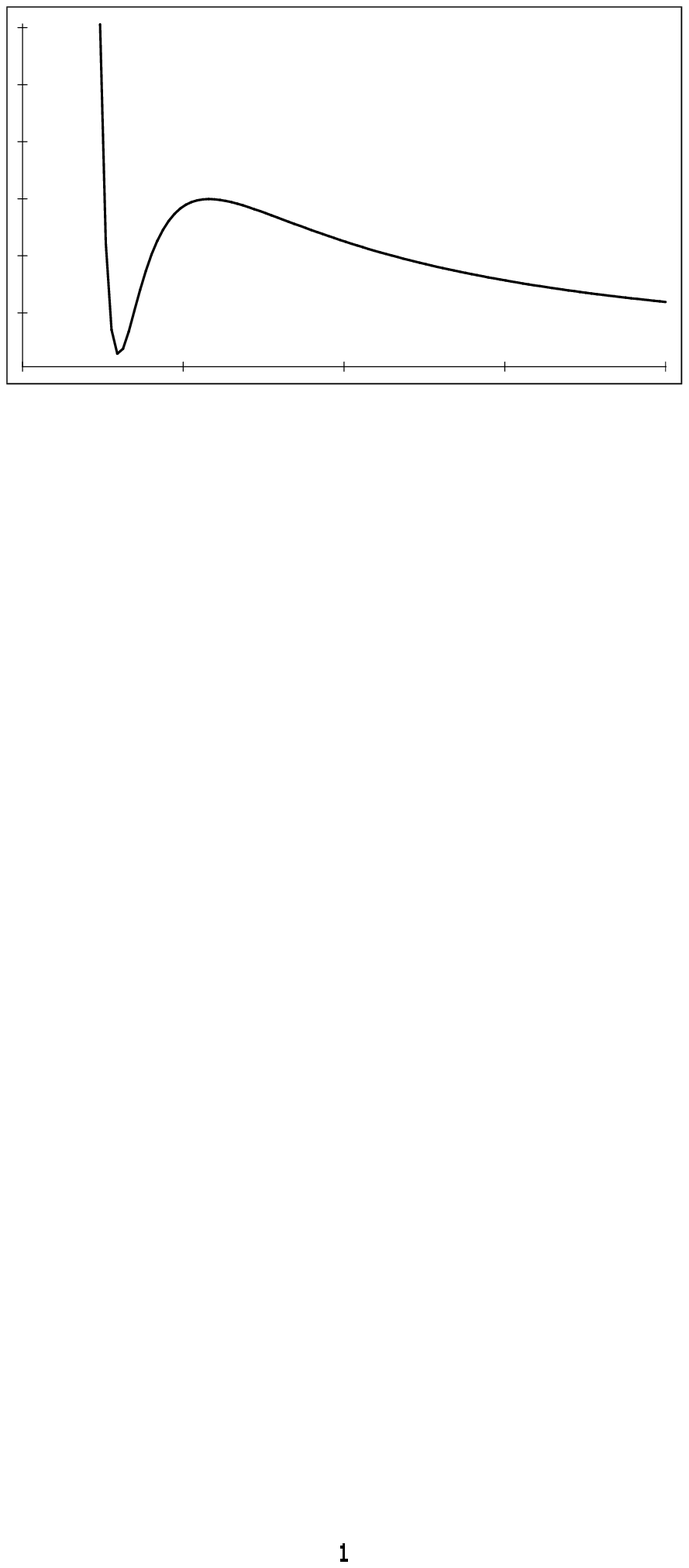}%
\\
Figure 1
\end{center}
%EndExpansion

\bigskip%

%TCIMACRO{\FRAME{dtbpFU}{3.5241in}{2.0704in}{0pt}{\Qcb{Figure 2}}{}%
%{fig2.eps}{\special{ language "Scientific Word";  type "GRAPHIC";
%width 3.5241in;  height 2.0704in;  depth 0pt;  original-width 3.4973in;
%original-height 8.0652in;  cropleft "0";  croptop "1";  cropright "1";
%cropbottom "0.7467";  filename '../Sb7/fig2.eps';file-properties "XNPEU";}}}%
%BeginExpansion
\begin{center}
\includegraphics[
trim=0.000000in 6.022285in 0.000000in 0.000000in,
height=2.0704in,
width=3.5241in
]%
{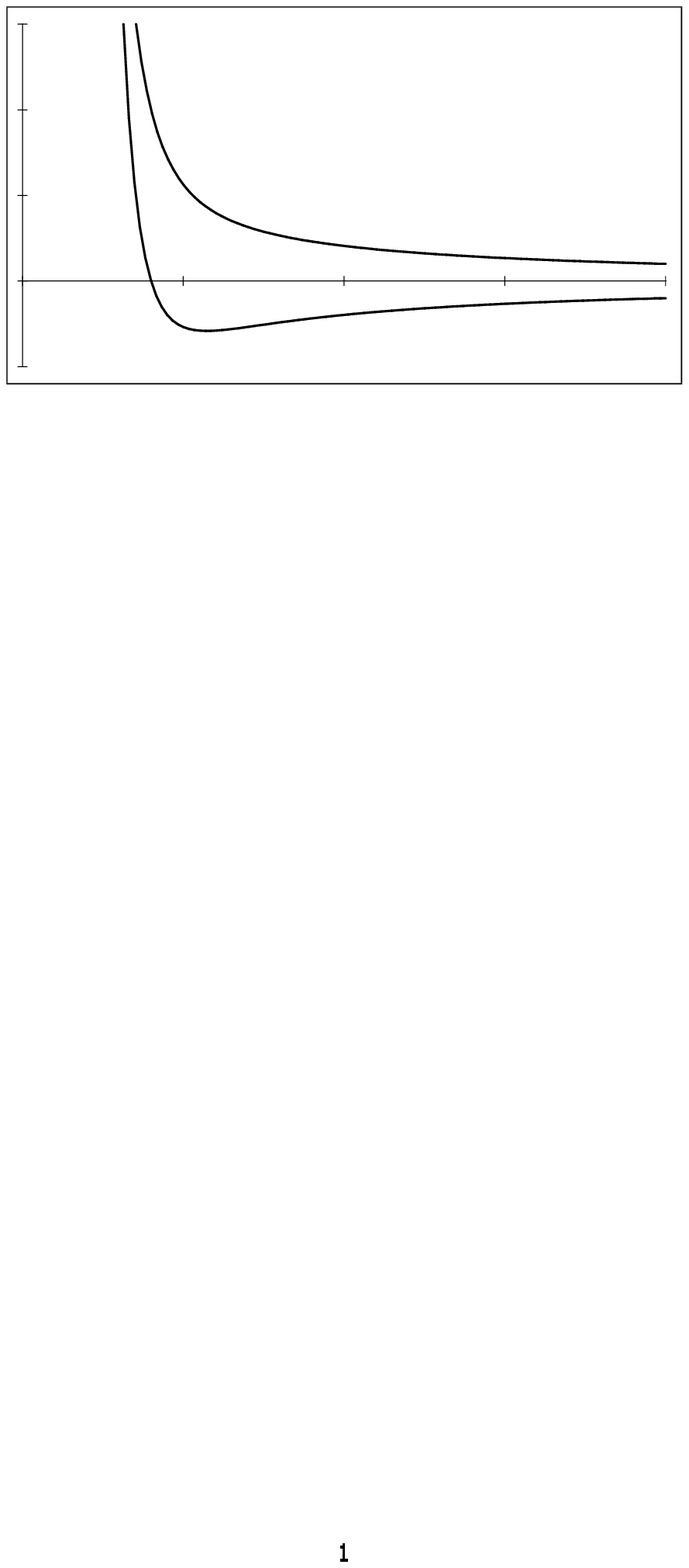}%
\\
Figure 2
\end{center}
%EndExpansion

\end{document}